\documentclass[conference,10pt]{IEEEtran}
\IEEEoverridecommandlockouts

\usepackage{makecell}
\usepackage{float}
\usepackage{amsmath,amssymb,amsfonts}
\usepackage{graphicx}
\usepackage{textcomp}
\usepackage{xcolor}
\usepackage{newtxtext}
\usepackage{newtxmath}
\usepackage{setspace}
\usepackage{algorithm}
\usepackage{algpseudocode}
\usepackage{url}

\def\BibTeX{{\rm B\kern-.05em{\sc i\kern-.025em b}\kern-.08em
T\kern-.1667em\lower.7ex\hbox{E}\kern-.125emX}}
\begin{document}

\title{Global Patterns in Student Stress and Academic Performance: A Machine Learning Study Using PISA 2022}

\author{\IEEEauthorblockN{Ani Ghazanchyan, Sachin Kumar${^*}$}
\IEEEauthorblockA{Javen P. and Sonia Akian College of Science \& Engineering, American University of Armenia, Yerevan, Armenia\\
ani\_ghazanchyan@edu.aua.am; s.kumar@aua.am \\ Corresponding author: Sachin Kumar (s.kumar@aua.am)}}

\maketitle

\begin{abstract}
Machine learning was applied to examine whether stress-related factors influence student performance in a consistent way across the world. The main goal of this project is to confirm or reject the existence of a similar global pattern by generalizing the findings that already exist in this field. We focused on various psychological indicators such as anxiety score, test anxiety, math anxiety, math confidence, wellbeing, and sense of belonging, along with several non-psychological factors for context. Machine learning was chosen due to the extremely large size of the PISA 2022 dataset and its ability to capture complex relationships that simpler methods may overlook. The analysis was conducted across six continents by splitting the dataset into six separate case studies. Feature engineering was performed manually for each region, while the same baseline models were trained to ensure a fair comparison. The results show that the negative effect of stress on performance is present and fairly consistent across all continents. Although some error remains—partly because stress is not the only factor shaping academic outcomes—the overall pattern is clear. Africa stood out as an outlier due to lower average educational and wellbeing levels and a higher proportion of missing data, yet even there the negative relationship remained observable.

\end{abstract}

\begin{IEEEkeywords}PISA 2022, Student performance, Stress factors, Cross-continental analysis, Psychological indicators, HistGradientBoosting.
\end{IEEEkeywords}

\section{Introduction and Background}
Assessing how stress and related psychological factors influence student performance is important because identifying consistent patterns can help educators reduce the negative effects of stress on learning.
When stress levels are high, students’ scores may reflect their emotional state rather than their true abilities. In this project, the focus is on variables such as math anxiety, self-reported math confidence, and other indicators that collectively represent students’ stress levels during assessment. Examining these factors together provides a clearer picture of how psychological pressure shapes performance outcomes. Machine learning is used because the PISA 2022 dataset\cite{oecd2022pisa}  (PISA: Programme for International Student Assessment) is extremely large and allows the application of powerful models. Powerful ML models can capture non-linear trends and produce accurate predictions, which simpler methods may overlook. This makes it possible to generalize findings that previously existed only at the level of individual schools, regions, or countries, and instead assess whether similar patterns appear across the world. Comparing continents is also meaningful due to cultural, geographic, and educational differences that may affect how stress influences performance. The main goal of this project is to evaluate the global impact of stress on academic outcomes and determine whether the effects are consistent across different world regions.

From a theoretical perspective, stress-related responses to academic evaluation are expected to exhibit both universal and culture-specific characteristics. Psychological stress activates common cognitive mechanisms that interfere with attention regulation and working memory, leading to performance inhibition across contexts \cite{eysenck2007anxiety}. At the same time, the way stress is experienced, expressed, and managed can vary substantially across cultural and educational environments. Cultural norms related to academic pressure, competitiveness, attitudes toward failure, and emotional expression influence how students perceive and respond to stressors in educational settings \cite{hofstede2001culture}. Moreover, large-scale assessments such as PISA rely heavily on self-reported measures of anxiety, well-being, and confidence, which are known to be affected by cross-cultural differences in response styles and self-evaluation tendencies \cite{heine2002self}\cite{oecd2018wellbeing}. As a result, while the overall direction of the stress–performance relationship is expected to remain negative across regions, its magnitude and the relative importance of specific stress-related factors may differ. This theoretical background motivates a continent-level analysis that allows both global consistency and regional variation in the stress–performance relationship to be examined simultaneously\cite{foley2017global,namkung2019meta,barroso2021meta}.

To contextualize this work, it is essential to position it within the growing body of international research exploring stress and academic outcomes. Recent studies using the PISA 2022 dataset have begun to quantify how psychological variables like anxiety and self-efficacy relate to performance across countries and cultures. The PISA 2022 dataset includes new measures related to student stress, anxiety, and self-efficacy, enabling global studies on psychological impacts on performance. Colomer et al.\ \cite{colomer2024mathanxiety} showed a consistent negative correlation between math anxiety and performance. Similar results were observed in ACER’s national report, where high stress resistance aligned with significantly better math scores in Australia \cite{acer2024stressgap}. Other research emphasized the positive role of self-confidence: Underwood \cite{underwood2025selfefficacy} and Niu et al.\ \cite{niu2025crosscultural} reported self-efficacy as a strong predictor of math achievement across several countries. Machine learning applications such as XGBoost and SHAP in Huang et al.\ \cite{huang2024mlshap} further confirmed these trends in East Asia. These studies support our analysis, which evaluates the global consistency of stress-performance effects using ML across continents \cite{zhang2019meta,barroso2021meta}. By identifying stress related psychological factors that show consistent associations with performance across continents, this study provides empirical grounding for the design and adaptation of educational psychology interventions in diverse international contexts.

\section{Dataset Description}

The analysis uses the PISA 2022 student dataset provided by the OECD \cite{oecd2022pisa}. This dataset captures a wide range of information about students, including academic performance, well-being indicators, psychological measures, social factors, family and peer relationships, study patterns, and many categorical survey responses.

\begin{table}[H]
\centering
\caption{Dataset sizes per continent after preprocessing.}
\label{tab:cont}
\begin{tabular}{lccccc}
\hline
\textbf{Cont.} & \textbf{Tot.} & \textbf{Tr.} & \textbf{Val.} & \textbf{Test} & \textbf{Feat.} \\
\hline
Europe         & 260821 & 182574 & 39123 & 39124 & 61 \\
N. America     & 67206  & 47044  & 10081 & 10081 & 55 \\
S. America     & 55871  & 39109  & 8381  & 8381  & 59 \\
Asia           & 197667 & 138366 & 29650 & 29651 & 58 \\
Africa         & 6867   & 4806   & 1030  & 1031  & 54 \\
Oceania        & 18119  & 12683  & 2718  & 2718  & 59 \\
\hline
\end{tabular}
\end{table}

Before cleaning, the dataset contained 613,744 samples and 1,248 features from 81 participating countries. Since our focus was specifically on stress-related factors, we reduced the dataset to roughly sixty relevant variables while keeping the full sample size unchanged. No countries were removed. We later dropped some additional features due to high proportions of missing values.

Data completeness varied substantially across continents, reflecting differences in survey participation, item nonresponse, and optional questionnaire modules. As a result, the proportion of missing values differed both across regions and across stress-related variables\cite{oecd2022vol2}.

We split the dataset geographically into six case studies representing six continents (Table~\ref{tab:cont}). The largest sample sizes were for Europe and Asia (around 200{,}000 each), the second largest was North America (\textasciitilde70{,}000), for South America the number of samples was \textasciitilde60{,}000, Oceania had \textasciitilde20{,}000, and Africa featured \textasciitilde10{,}000 being the smallest among all datasets we will use. Africa also yielded the fewest usable features because of higher missing-value rates.

\begin{table*}[t]
\centering
\caption{Top stress- and well-being--related features used in the models. Selection based on conceptual relevance to stress and global SHAP importance across continents.}
\label{tab:stress_feature_summary}
\resizebox{\textwidth}{!}{%
\begin{tabular}{llll}
\hline
\textbf{Feature} & \textbf{Type} & \textbf{Construct} & \textbf{Short Description} \\
\hline
ANXMAT     & Numerical index      & Math anxiety          & Level of math-related anxiety \\
PSYCHSYM   & Numerical index      & Psychosomatic symptoms& Symptoms such as headaches, stomach pain \\
STRESAGR   & Numerical index      & Stress resistance     & Perceived ability to cope with stress \\
WB154Q04HA & Ordinal frequency    & Depressive affect     & Frequency of feeling depressed \\
WB154Q05HA & Ordinal frequency    & Irritability          & Feeling irritable or bad tempered \\
WB154Q06HA & Ordinal frequency    & Nervousness           & Frequency of feeling nervous or tense \\
WB154Q07HA & Ordinal frequency    & Sleep problems        & Difficulty falling asleep \\
FEELSAFE   & Numerical index      & School safety         & Perceived safety in school environment \\
BULLIED    & Numerical index      & Bullying/victimization& Frequency of being bullied \\
RELATST    & Numerical index      & Teacher support       & Perceived student--teacher relationship \\
BELONG     & Numerical index      & School belonging      & Sense of belonging at school \\
LIFESAT    & Numerical index      & Life satisfaction     & Overall life satisfaction \\
FAMSUPSL   & Numerical index      & Family academic support & Support for learning at home \\
EMOCOAGR   & Numerical index      & Emotion regulation    & Staying calm when under pressure \\
SKIPPING   & Numerical (count)    & School avoidance      & Skipping school or classes \\
EXERPRAC   & Numerical (frequency)& Coping behavior       & Frequency of exercising \\
ICTWKDY    & Numerical (time)     & Daily screen time     & ICT use during weekdays \\
MATHEFF    & Numerical index      & Math self-efficacy    & Confidence solving math tasks \\
HOMEPOS    & Numerical index      & Home possessions      & Access to books, internet, etc. \\
ESCS       & Numerical index      & Socioeconomic status  & Combined economic, social, cultural status \\
\hline
\end{tabular}%
}
\end{table*}

Table~\ref{tab:stress_feature_summary} summarizes the key variables used in the analysis, focusing on stress- and well-being--related features selected based on theoretical relevance and SHAP-based model importance.

The preprocessing steps were to handle missing values, construct the target variable, scale the features, remove variables directly related to performance, and create train/validation/test splits. Missing data were addressed in two stages. First, columns with very high missing-value percentages were dropped, with thresholds chosen manually for each continent based on sample size and importance. Second, remaining missing values were imputed. KNN imputation was initially attempted, but was computationally infeasible because of hardware limitations for very large datasets like Europe and Asia, so we used scikit-learn's \texttt{SimpleImputer} with the median strategy instead.

The target variable was engineered by averaging students’ available exam scores across subjects to reduce noise and bias. Next step was to normalize continuous features using mean and standard deviation. The splitting of the data was done as follows: 70\% training, 15\% validation, and 15\% testing sets. After these steps, each continental dataset was ready for modeling.

\section{Proposed Methodology}
\subsection{Handling Missing Data and Feature Selection}

Because of great variation in data completeness across continents, a uniform missing-value threshold would have disproportionately reduced the usable sample size in regions with higher item nonresponse. 
Instead, continent-specific thresholds were applied following a 
consistent rule: features were retained if their missing-value proportion 
allowed stable imputation without reducing the effective sample size 
below a meaningful level for modeling. 

Thresholds were therefore selected by jointly considering (i) the relative 
missingness of each feature within a continent, (ii) the total sample size 
available, and (iii) the theoretical relevance of the feature to stress 
and well-being. This procedure ensured that regions with smaller datasets 
were not unfairly penalized while preserving comparability across continents.

To verify that results were not overly sensitive to these thresholds, 
we observed that the ordering of model performance and the direction of 
stress-related effects remained stable across continents, despite minor 
differences in feature availability. This indicates that the core findings 
are robust to reasonable variations in preprocessing decisions.

\subsection{Algorithms used}

In this project, three baseline machine learning models were used: Linear Regression, Random Forest, and HistGradientBoosting. These models were selected because they provide a balanced comparison between simple, interpretable methods and more powerful non-linear approaches. Linear Regression was included as a baseline due to its simplicity and interpretability. Mathematically, Linear Regression models the predicted performance as
\[
\hat{y} = X\beta,
\]
where the parameters are estimated by minimizing the squared loss
\[
\beta^{*} = \arg\min_{\beta} \| y - X\beta \|_{2}^{2}.
\]

Random Forest was chosen because it can capture complex non-linear relationships while remaining relatively robust and easy to use. A Random Forest prediction is computed as the average over \(T\) decision trees:
\[
\hat{y}(x) = \frac{1}{T} \sum_{t=1}^{T} \hat{y}_{t}(x).
\]

HistGradientBoosting was used as the main high-performance model, as it is specifically optimized for large tabular datasets such as PISA. Gradient boosting builds the model iteratively. At each boosting step \(t\), the model is updated according to
\[
f_{t}(x) = f_{t-1}(x) + \eta\, h_t(x),
\]
where \(h_t(x)\) is a regression tree fitted to the negative gradient of the loss function, and \(\eta\) is the learning rate. For regression, the optimized loss function is the mean squared error:
\[
L = \frac{1}{n} \sum_{i=1}^{n} (y_i - f(x_i))^{2}.
\]

Heavy hyperparameter tuning was not performed because of the extremely large size of the dataset, which makes extended tuning computationally impractical. Linear Regression requires no hyperparameter adjustments. For Random Forest, only light modifications were made—such as reducing the number of trees—to avoid long training times. In particular, the number of trees was reduced from the typical 150–200 range to 80 estimators, and the maximum depth was restricted, ensuring that training remained fast on the large PISA dataset while still capturing essential non-linear patterns.
Each model used its default parameters unless otherwise noted (as in the Random Forest adjustments), meaning no exhaustive hyperparameter search was performed due to the dataset’s scale.
Concerning HistGradientBoosting, we mostly used its default settings, as the algorithm is already designed for high efficiency and scalability on datasets with hundreds of thousands of rows.

Other models such as XGBoost, LightGBM, or neural networks were considered but ultimately not used. Many either demand substantially more computational resources or careful tuning (which was impractical given our data size), or they risked overfitting. 

The final choice of Linear Regression, Random Forest, and HistGradientBoosting, however, reflects a trade-off between interpretability, robustness, and computational feasibility.

\subsection{Evaluation Metric}
Three metrics were used for the evaluation of the performance of our model: Mean Absolute Error (MAE), Root Mean Squared Error (RMSE), and the R² score. The perspectives these metrics provide on predictive accuracy and error behavior are complementary. R² shows how much of the target variable's variance is explained by the model, making it useful for understanding overall fit. RMSE penalizes larger errors, which highlights cases where the model makes significant mistakes. MAE, on the other hand, is easier to interpret and more robust to outliers, offering a stable estimate of average prediction error.

To formalize these evaluation metrics in the context of our problem—predicting student performance scores ($\hat{y}_i$) from stress-related features in the PISA 2022 dataset—let $y_i$ be the true observed score for student $i$, for $i = 1, 2, \dots, n$. Then, the metrics are defined as follows:

\begin{equation}
\text{MAE} = \frac{1}{n} \sum_{i=1}^{n} \left| y_i - \hat{y}_i \right|
\end{equation}

\begin{equation}
\text{RMSE} = \sqrt{ \frac{1}{n} \sum_{i=1}^{n} (y_i - \hat{y}_i)^2 }
\end{equation}

\begin{equation}
R^2 = 1 - \frac{ \sum_{i=1}^{n} (y_i - \hat{y}_i)^2 }{ \sum_{i=1}^{n} (y_i - \bar{y})^2 }, \quad \text{where} \quad \bar{y} = \frac{1}{n} \sum_{i=1}^{n} y_i
\end{equation}

MAE estimates the average magnitude of prediction errors, RMSE emphasizes larger errors due to squaring, and $R^2$ measures the proportion of variance in student scores that is captured by the model. These metrics provide complementary perspectives when evaluating model performance across validation and test sets in each continent.

We didn't proceed with MSE as we had already computed RMSE, which provides the same information in the original unit scale of the target variable. That makes it easier to interpret. Adjusted R² was also not used because it does not have great benefit for datasets of large sizes; with hundreds of thousands of samples, the adjustment term becomes negligible and does not change the interpretation meaningfully.

For completeness, the Mean Squared Error (MSE) and Adjusted $R^2$ are given below:
\begin{equation}
\text{MSE} = \frac{1}{n} \sum_{i=1}^{n} (y_i - \hat{y}_i)^2
\end{equation}

\begin{equation}
R^2_{\text{adj}} = 1 - \left( \frac{1 - R^2}{n - p - 1} \right) (n - 1)
\end{equation}

Here, $n$ is the number of observations and $p$ is the number of predictors. MSE is functionally equivalent to RMSE but lacks interpretability in the original units. Adjusted $R^2$ introduces a penalty for adding predictors but becomes almost identical to $R^2$ in large datasets like ours, where $n \gg p$, rendering its correction term negligible.

The results obtained were used to access the consistency of each model's performance. Based on these metrics, we selected the strongest model for each continent for further analysis and cross-continental comparison.

\section{Experiments and Results}
\subsection{Model Performance Across Continents}
The three baseline models—Linear Regression, Random Forest, and HistGradientBoosting—were trained and evaluated separately on each continent, producing a total of eighteen results. Validation performance was used to select the best model per region, and HistGradientBoosting performed best across all six continents. Although each region had slightly different feature sets due to missing values and manual preprocessing decisions, the same train/validation/test split of 70/15/15 was used everywhere, which made comparisons fair and consistent.

\begin{figure}[H]
    \centering
    \includegraphics[width=0.46\textwidth]{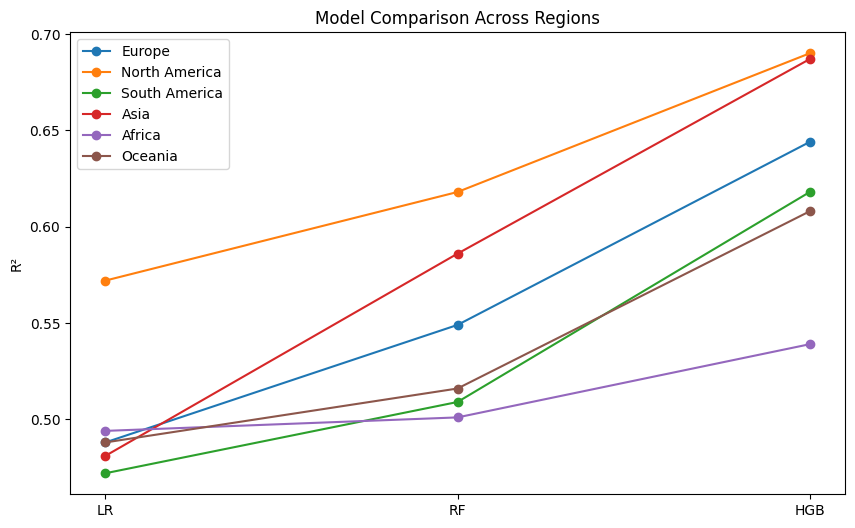}
    \caption{The R² score comparison of Linear Regression (LR), Random Forest (RF), and HistGradientBoosting (HGB) across six continents regions. HGB consistently outperforms the other models on all continents, while LR shows the weakest performance.}
    \label{fig:model_comparison}
\end{figure}

Across continents, clear performance patterns emerged. North America and Asia achieved the strongest results, with the highest R² values on both train and test sets. Europe and South America showed moderate results, while Oceania performed slightly worse, with noticeably higher error. For Africa, we had the lowest R² and greatest instability consistently. This aligns with its smaller sample size and higher proportion of missing or dropped features. Despite these regional differences, the ordering of continents remained very similar across train and test results, showing stable behavior.

Model rankings were also consistent. HistGradientBoosting was the strongest model in every continent. Random Forest performed second across all regions, while Linear Regression consistently had the weakest results. The performance gap between models was larger in regions with more data, while in Africa, Random Forest and HistGradientBoosting behaved more similarly.

Error metrics further supported these trends. South America had the lowest MAE, while Oceania exhibited the highest error. Africa also showed low MAE, but this is likely because the overall variance of exam scores in that region is smaller, not because the model fit well. Europe, Asia, and North America all showed mid-range MAE values. Importantly, MAE did not correlate perfectly with R²: Africa had one of the lowest MAE scores but also the lowest R², highlighting that lower absolute error does not necessarily imply a strong predictive model.

Most importantly, the stress–performance effect remained consistently negative across all continents, in both train and test sets. This implies that the models did successfully capture whatever negative relationship existed in the data. Minor fluctuations in R² indicate the results are stable on average. Africa had the weakest correlation (test), influenced by its smaller dataset and feature group, yet the effect was still present. These results show that the negative impact of stress-related factors on performance persists globally despite cultural, geographic, and educational differences. \cite{foley2017global}

\begin{figure}[ht]
    \centering
    \includegraphics[width=0.43\textwidth]{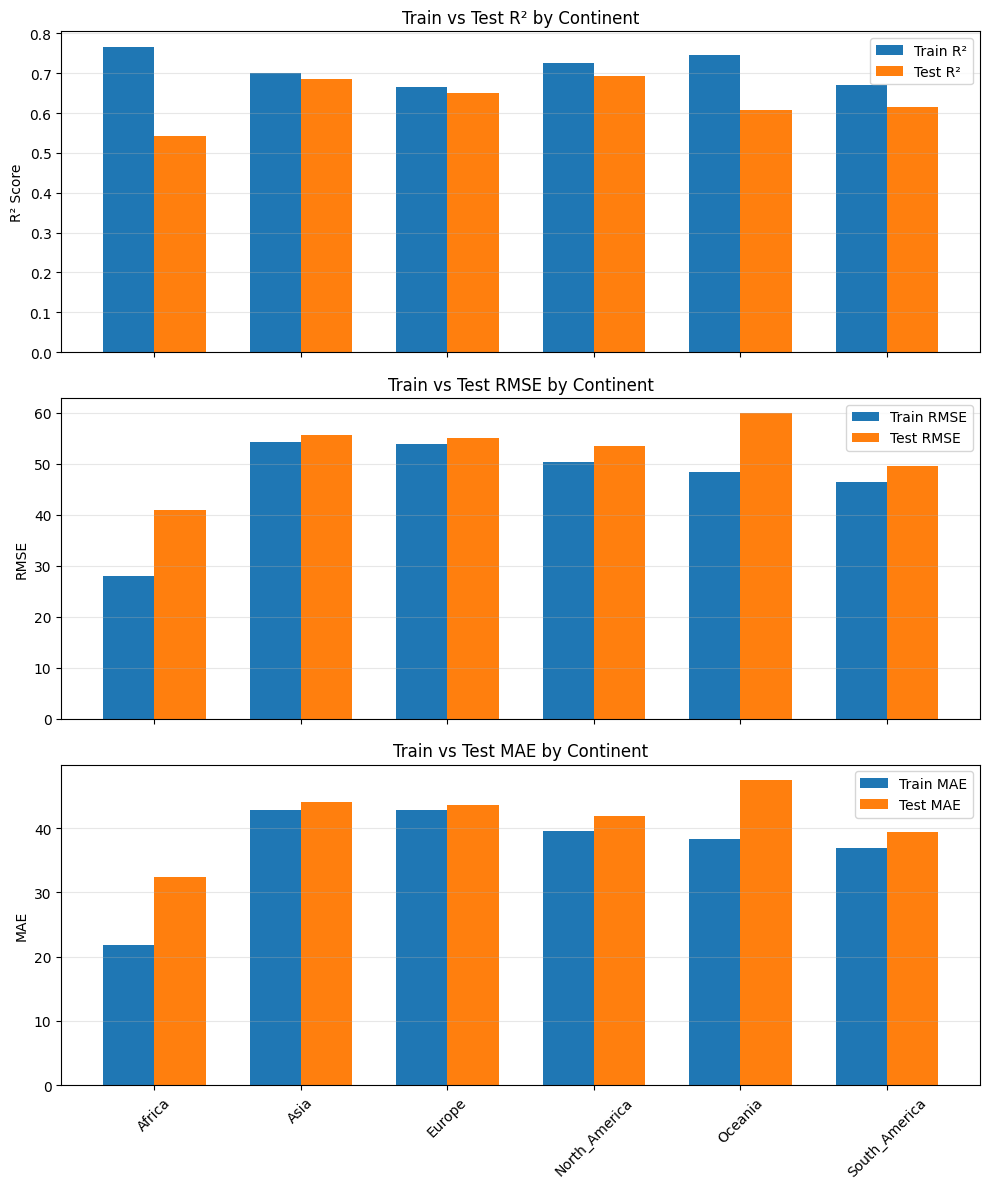}
    \caption{Comparison of train and test performance across continents for HistGradientBoosting. 
    The top subplot shows R\textsuperscript{2} values. The middle subplot shows RMSE while the bottom one reports MAE values, which 
    follow a similar pattern with only minor fluctuations between the two splits.}
    \label{fig:val_test_comparison}
\end{figure}

\begin{figure}[H]
    \centering
    \includegraphics[width=0.4\textwidth]{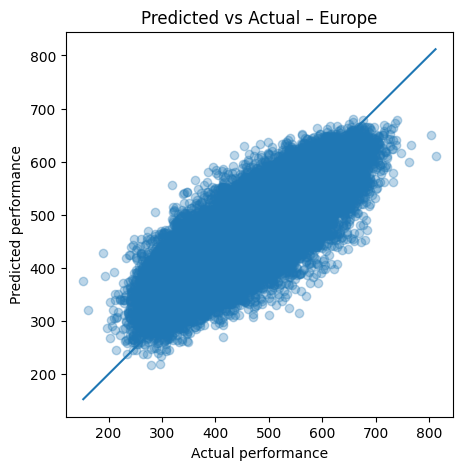}
    \caption{Predicted vs actual student performance for Europe using the HistGradientBoosting model. The tight clustering along the diagonal line indicates strong model stability and a well-captured underlying trend.}
    \label{fig:pred_vs_actual_europe}
\end{figure}

Overall, the results provide strong evidence that stress and related psychological factors negatively affect student performance worldwide, and that machine learning models can consistently detect this pattern across distinct regions.

\subsection{Quantifying and Interpreting the Stress--Performance Effect}

To quantify the magnitude and structure of the stress--performance relationship, we analyzed model behavior using SHAP (SHapley Additive exPlanations). SHAP values decompose each prediction into additive feature contributions, allowing both the direction and relative strength of each variable’s effect on predicted student performance to be examined.

Across all six continents, stress-related variables consistently exert a negative influence on predicted performance. Higher mathematics anxiety, school disengagement (e.g., skipping or repetition), social stress (e.g., bullying), and weaker emotional or institutional support are associated with negative SHAP values, lowering predicted scores. In contrast, protective factors such as mathematics self-efficacy, home educational resources, family support, and coping behaviors (e.g., exercise) consistently contribute positive SHAP values.\cite{oecd2018wellbeing}

\begin{figure}[ht]
    \centering
    \includegraphics[width=1\linewidth]{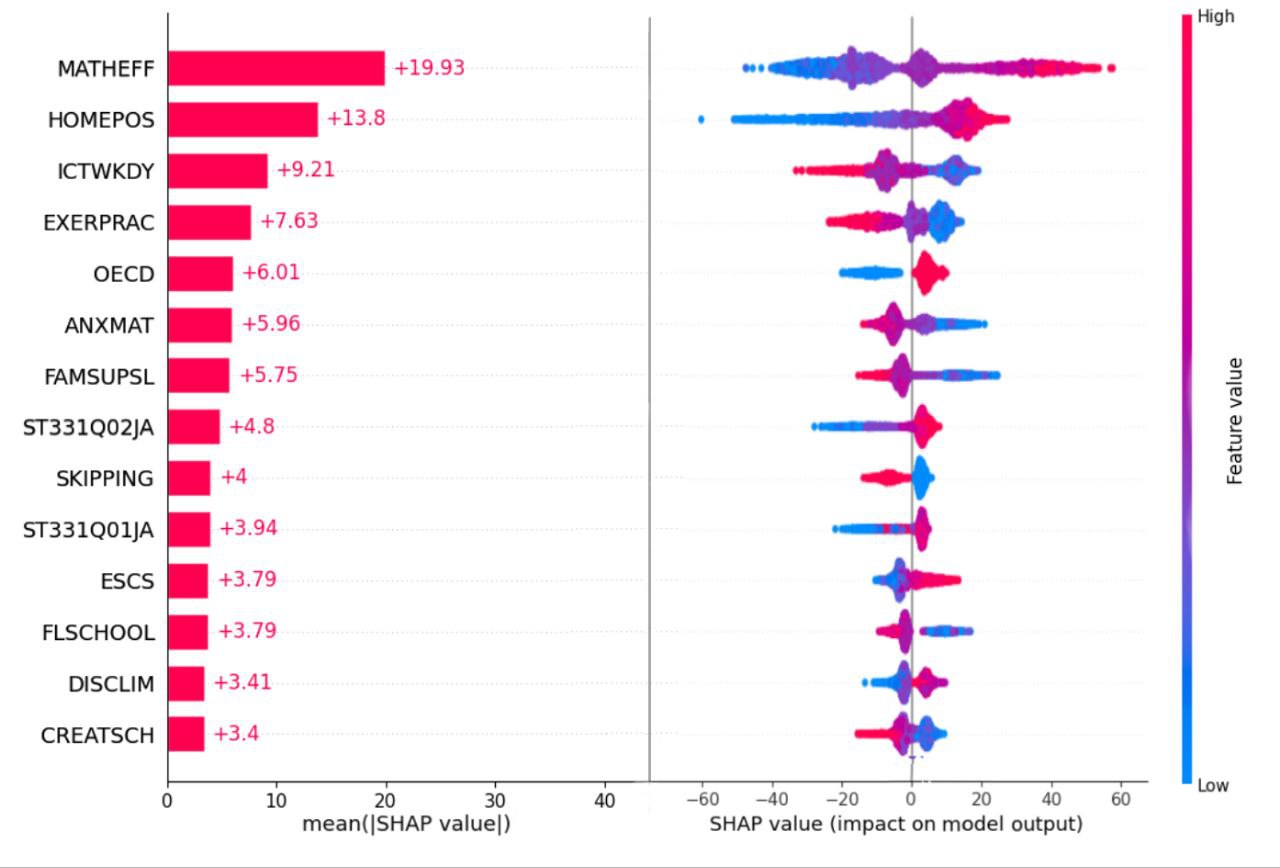}
    \caption{SHAP summary plot for Europe using the HistGradientBoosting model. Stress-related features such as math anxiety, skipping behavior, and family support show strong influence on predicted performance, with higher anxiety and disengagement associated with lower predicted scores.}
    \label{fig:shap_europe}
\end{figure}

Europe exhibits the most stable and interpretable pattern. As shown in Figure~\ref{fig:shap_europe}, mathematics self-efficacy and home resources dominate positive contributions, while anxiety and disengagement strongly reduce predicted outcomes. North America and South America display closely related structures, although formal academic indicators such as GRADE play a more prominent role in South America, suggesting partial mediation of stress effects through institutional performance markers.

In Asia, self-efficacy, home resources, and effort-related variables dominate feature importance, while stress indicators remain present but are relatively less influential. This suggests that performance-oriented factors partially buffer the observable impact of stress without eliminating it. Oceania follows a pattern similar to Europe and North America, with consistent directional effects but greater dispersion, reflecting its smaller sample size rather than a distinct underlying mechanism.

Africa presents the most distinct structure. GRADE emerges as the dominant predictor, reflecting the strong influence of institutional and structural educational factors. Nevertheless, stress-related variables such as anxiety, bullying, weak school belonging, and disengagement behaviors continue to exert negative effects. These effects are more scattered and less stable, consistent with higher missing-data rates, greater heterogeneity, and smaller sample size rather than a fundamentally different stress--performance relationship.

Overall, the SHAP analysis demonstrates that the stress--performance relationship is not only globally present but also substantial in magnitude. While the relative importance of individual variables varies across regions, the direction and structure of stress effects remain remarkably consistent. Regional differences primarily reflect data stability, socioeconomic context, and educational structure rather than qualitative differences in how stress influences academic performance.

\subsection{Quantifying and Interpreting the Stress--Performance Effect}

To quantify how stress-related and contextual variables influence predicted student performance, 
SHAP analysis was applied to the HistGradientBoosting models. SHAP values indicate both the magnitude 
and direction of each feature’s contribution to the model output, allowing a detailed examination 
of how stress-related factors affect predictions across regions.

Due to space limitations, SHAP summary plots are shown for Europe and Africa only. Europe represents 
the largest and most stable dataset, providing a clear reference pattern under favorable data conditions, 
while Africa represents the most challenging case, with a smaller sample size and higher missing-data 
rates. The remaining continents (North America, South America, Asia, and Oceania) exhibit feature 
importance patterns that closely resemble those observed in Europe. These plots were analyzed in full 
but are omitted here for brevity.

As shown in Fig.~4, the SHAP summary plot for Europe reveals that math self-efficacy (MATHEFF), home 
educational resources (HOMEPOS), and family academic support (FAMSUPSL) contribute positively to 
predicted performance, while higher math anxiety (ANXMAT), skipping behavior (SKIPPING), and weaker 
engagement tend to push predictions downward. The color gradients further indicate that higher anxiety 
levels and disengagement behaviors consistently correspond to negative SHAP values, reinforcing the 
negative stress--performance relationship observed in the predictive results.

\begin{figure}[ht]
    \centering
    \includegraphics[width=1\linewidth]{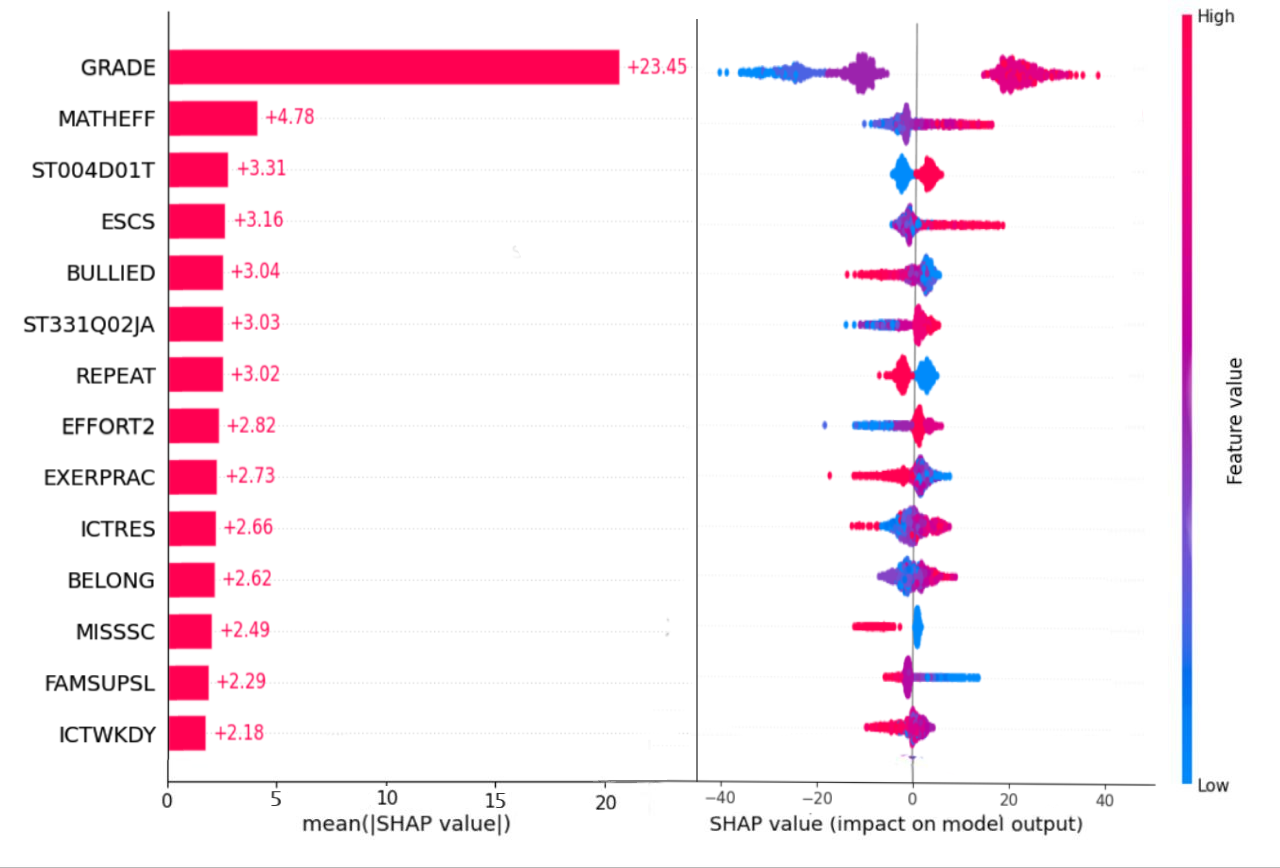}
    \caption{SHAP summary plot for Africa using the HistGradientBoosting model. While many of the same stress-related features remain influential, the effects in Africa are more scattered and less stable due to the smaller sample size and higher missing data. GRADE appeared as the strongest predictor, but higher anxiety, skipping behavior, and weaker support still lowered predicted performance, consistent with global patterns.}
    \label{fig:shap_africa}
\end{figure}

In contrast, Fig.~5 shows the SHAP summary plot for Africa, where the strongest predictor is GRADE, 
a school-reported indicator of overall academic standing. Despite this difference, many of the same 
psychological and behavioral variables remain influential. Higher math anxiety, lower school belonging 
(BELONG), and increased disengagement behaviors still contribute negatively to performance predictions, 
although the effects are more scattered and less stable. This increased variability is consistent with 
Africa’s smaller sample size and higher data sparsity.

Taken together, the SHAP analyses demonstrate that while the relative importance of specific predictors 
varies across regions, the overall direction of stress-related effects remains consistent. Higher levels 
of stress and disengagement are associated with lower predicted performance, whereas self-efficacy, 
support, and resource-related factors act as protective influences across continents. Moreover, 
stress-related variables such as mathematics anxiety, school disengagement, and sense of belonging 
consistently rank among the top contributors by mean absolute SHAP value, indicating that their impact 
on predicted performance is comparable in magnitude to major socioeconomic and home-resource factors. 
These results confirm that stress-related mechanisms play a substantial and globally relevant role in shaping student academic outcomes.

\section{Conclusion}
The results of this study show a consistent negative relationship between stress-related factors and student performance across all examined regions. Even with differences in data availability and the need for manual processing in each continental subset, the overall pattern remained stable, allowing the effect to be generalized globally. While varying feature sets and uneven sample sizes introduce some error and limit direct comparability across continents, model performance was sufficiently strong to reliably identify the core relationship between stress indicators and academic outcomes. It should be noted that this analysis is correlational; while stress indicators align with poorer performance, experimental or longitudinal studies would be needed to confirm causal effects.

Several limitations must therefore be acknowledged. First, the observational nature of the PISA 2022 dataset prevents causal interpretation of the reported relationships. It is possible that lower academic performance contributes to higher reported stress, or that both outcomes are jointly influenced by unobserved factors.\cite{carey2016direction} Second, although the models include socioeconomic status and home-resource indicators, residual confounding remains likely. Differences in educational systems, curriculum difficulty, grading practices, and access to academic support may simultaneously affect both stress levels and performance outcomes across regions. Third, stress- and well-being–related variables rely on self-reported survey responses, which may be influenced by cultural norms, language interpretation, and reporting tendencies. Such cross-cultural variation in self-reporting may partially explain the increased instability observed in regions with smaller sample sizes or higher missingness, particularly Africa.

Despite these limitations, the consistency of the negative stress–performance relationship across six continents, multiple evaluation metrics, and independent train and test splits suggests that the findings reflect a robust global pattern rather than region-specific artifacts. From a practical perspective, the results highlight the importance of addressing student stress as a component of educational policy and assessment design. Interventions aimed at reducing anxiety, strengthening self-efficacy, and improving emotional and social support structures may help ensure that academic evaluations better reflect students’ knowledge and abilities rather than their psychological state.\cite{yeager2019mindset} Future work could build on this analysis by examining which institutional or behavioral factors most effectively reduce stress and by exploring causal mechanisms through longitudinal or experimental study designs.








\section*{Conflicts of interest} The authors declare no conflicts of interest.

\end{document}